\documentstyle[twoside]{article}

\catcode`\@=11
\long\def\@makefntext#1{
\protect\noindent \hbox to 3.2pt {\hskip-.9pt  
$^{{\eightrm\@thefnmark}}$\hfil}#1\hfill}		%CAN BE USED 

\def\thefootnote{\fnsymbol{footnote}}
\def\@makefnmark{\hbox to 0pt{$^{\@thefnmark}$\hss}}	%ORIGINAL 
	
\def\ps@myheadings{\let\@mkboth\@gobbletwo
\def\@oddhead{\hbox{}
\rightmark\hfil\eightrm\thepage}   
\def\@oddfoot{}\def\@evenhead{\eightrm\thepage\hfil
\leftmark\hbox{}}\def\@evenfoot{}
\def\sectionmark##1{}\def\subsectionmark##1{}}

\oddsidemargin=\evensidemargin
\renewcommand{\thefootnote}{\fnsymbol{footnote}}

\newcounter{sectionc}\newcounter{subsectionc}\newcounter{subsubsectionc}
\renewcommand{\section}[1] {\vspace{12pt}\addtocounter{sectionc}{1} 
\setcounter{subsectionc}{0}\setcounter{subsubsectionc}{0}\noindent 
	{\tenbf\thesectionc. #1}\par\vspace{5pt}}
\renewcommand{\subsection}[1] {\vspace{12pt}\addtocounter{subsectionc}{1} 
	\setcounter{subsubsectionc}{0}\noindent 
	{\bf\thesectionc.\thesubsectionc. {\kern1pt \bfit #1}}\par\vspace{5pt}}
\renewcommand{\subsubsection}[1] {\vspace{12pt}\addtocounter{subsubsectionc}{1}
	\noindent{\tenrm\thesectionc.\thesubsectionc.\thesubsubsectionc.
	{\kern1pt \tenit #1}}\par\vspace{5pt}}
\newcommand{\nonumsection}[1] {\vspace{12pt}\noindent{\tenbf #1}
	\par\vspace{5pt}}

\newcounter{appendixc}
\newcounter{subappendixc}[appendixc]
\newcounter{subsubappendixc}[subappendixc]
\renewcommand{\thesubappendixc}{\Alph{appendixc}.\arabic{subappendixc}}
\renewcommand{\thesubsubappendixc}
	{\Alph{appendixc}.\arabic{subappendixc}.\arabic{subsubappendixc}}

\renewcommand{\appendix}[1] {\vspace{12pt}
        \refstepcounter{appendixc}
        \setcounter{figure}{0}
        \setcounter{table}{0}
        \setcounter{lemma}{0}
        \setcounter{theorem}{0}
        \setcounter{corollary}{0}
        \setcounter{definition}{0}
        \setcounter{equation}{0}
        \renewcommand{\thefigure}{\Alph{appendixc}.\arabic{figure}}
        \renewcommand{\thetable}{\Alph{appendixc}.\arabic{table}}
        \renewcommand{\theappendixc}{\Alph{appendixc}}
        \renewcommand{\thelemma}{\Alph{appendixc}.\arabic{lemma}}
        \renewcommand{\thetheorem}{\Alph{appendixc}.\arabic{theorem}}
        \renewcommand{\thedefinition}{\Alph{appendixc}.\arabic{definition}}
        \renewcommand{\thecorollary}{\Alph{appendixc}.\arabic{corollary}}
        \renewcommand{\theequation}{\Alph{appendixc}.\arabic{equation}}
        \noindent{\tenbf Appendix \theappendixc #1}\par\vspace{5pt}}
\newcommand{\subappendix}[1] {\vspace{12pt}
        \refstepcounter{subappendixc}
        \noindent{\bf Appendix \thesubappendixc. {\kern1pt \bfit #1}}
	\par\vspace{5pt}}
\newcommand{\subsubappendix}[1] {\vspace{12pt}
        \refstepcounter{subsubappendixc}
        \noindent{\rm Appendix \thesubsubappendixc. {\kern1pt \tenit #1}}
	\par\vspace{5pt}}

\newcommand{\textlineskip}{\baselineskip=13pt}
\newcommand{\smalllineskip}{\baselineskip=10pt}

\def\eightcirc{
\begin{picture}(0,0)
\put(4.4,1.8){\circle{6.5}}
\end{picture}}
\def\eightcopyright{\eightcirc\kern2.7pt\hbox{\eightrm c}} 

\newcommand{\copyrightheading}[1]
	{\vspace*{-2.5cm}\smalllineskip{\flushleft
	{\footnotesize International Journal of Modern Physics A, #1}\\
	{\footnotesize $\eightcopyright$\, World Scientific Publishing
	 Company}\\
	 }}

\newcommand{\publisher}[2]{{\begin{center}\footnotesize\smalllineskip 
	Received #1\\
	Revised #2
	\end{center}
	}}

\def\abstracts#1#2#3{{
	\centering{\begin{minipage}{4.5in}\baselineskip=10pt\footnotesize
	\parindent=0pt #1\par 
	\parindent=15pt #2\par
	\parindent=15pt #3
	\end{minipage}}\par}} 

\renewenvironment{thebibliography}[1]
	{\frenchspacing
	 \ninerm\baselineskip=11pt
	 \begin{list}{\arabic{enumi}.}
	{\usecounter{enumi}\setlength{\parsep}{0pt}
	 \setlength{\leftmargin 12.7pt}{\rightmargin 0pt} %FOR 1--9 ITEMS
	 \setlength{\itemsep}{0pt} \settowidth
	{\labelwidth}{#1.}\sloppy}}{\end{list}}

\newcounter{itemlistc}
\newcounter{romanlistc}
\newcounter{alphlistc}
\newcounter{arabiclistc}

\newcommand{\fcaption}[1]{
        \refstepcounter{figure}
        \setbox\@tempboxa = \hbox{\footnotesize Fig.~\thefigure. #1}
        \ifdim \wd\@tempboxa > 5in
           {\begin{center}
        \parbox{5in}{\footnotesize\smalllineskip Fig.~\thefigure. #1}
            \end{center}}
        \else
             {\begin{center}
             {\footnotesize Fig.~\thefigure. #1}
              \end{center}}
        \fi}

\newcommand{\tcaption}[1]{
        \refstepcounter{table}
        \setbox\@tempboxa = \hbox{\footnotesize Table~\thetable. #1}
        \ifdim \wd\@tempboxa > 5in
           {\begin{center}
        \parbox{5in}{\footnotesize\smalllineskip Table~\thetable. #1}
            \end{center}}
        \else
             {\begin{center}
             {\footnotesize Table~\thetable. #1}
              \end{center}}
        \fi}

\def\@citex[#1]#2{\if@filesw\immediate\write\@auxout
	{\string\citation{#2}}\fi
\def\@citea{}\@cite{\@for\@citeb:=#2\do
	{\@citea\def\@citea{,}\@ifundefined
	{b@\@citeb}{{\bf ?}\@warning
	{Citation `\@citeb' on page \thepage \space undefined}}
	{\csname b@\@citeb\endcsname}}}{#1}}

\newif\if@cghi
\def\cite{\@cghitrue\@ifnextchar [{\@tempswatrue
	\@citex}{\@tempswafalse\@citex[]}}
\def\citelow{\@cghifalse\@ifnextchar [{\@tempswatrue
	\@citex}{\@tempswafalse\@citex[]}}
\def\@cite#1#2{{$\null^{#1}$\if@tempswa\typeout
	{IJCGA warning: optional citation argument 
	ignored: `#2'} \fi}}

\def\pmb#1{\setbox0=\hbox{#1}
	\kern-.025em\copy0\kern-\wd0
	\kern.05em\copy0\kern-\wd0
	\kern-.025em\raise.0433em\box0}

\def\fnt#1#2{\footnotetext{\kern-.3em
	{$^{\mbox{\scriptsize #1}}$}{#2}}}

\def\fpage#1{\begingroup
\voffset=.3in
\thispagestyle{empty}\begin{table}[b]\centerline{\footnotesize #1}
	\end{table}\endgroup}

\def\runninghead#1#2{\pagestyle{myheadings}
\markboth{{\protect\footnotesize\it{\quad #1}}\hfill}
{\hfill{\protect\footnotesize\it{#2\quad}}}}
\font\tenrm=cmr10
\font\tenit=cmti10 
\font\tenbf=cmbx10
\font\bfit=cmbxti10 at 10pt
\font\ninerm=cmr9

\font\eightrm=cmr8

\def\qed{\hbox{${\vcenter{\vbox{			%HOLLOW SQUARE
   \hrule height 0.4pt\hbox{\vrule width 0.4pt height 6pt
   \kern5pt\vrule width 0.4pt}\hrule height 0.4pt}}}$}}

\renewcommand{\thefootnote}{\fnsymbol{footnote}}	%USE SYMBOLIC FOOTNOTE

\begin{document}
\input epsf

\runninghead{Quarkonium Production at $Z^0$ and in $\Upsilon(1S)$ Decay} 
{Quarkonium Production at $Z^0$ and in $\Upsilon(1S)$ Decay}

\normalsize\textlineskip
\thispagestyle{empty}
\setcounter{page}{1}

\copyrightheading{}			%{Vol. 0, No. 0 (1993) 000--000}

\vspace*{0.88truein}

\fpage{1}
\centerline{\bf QUARKONIUM PRODUCTION AT $Z^0$ AND IN $\Upsilon(1S)$ DECAY}
%\vspace*{0.035truein}
%\centerline{\bf MANUSCRIPTS USING COMPUTER SOFTWARE}
\vspace*{0.37truein}
\centerline{\footnotesize KINGMAN CHEUNG}
\vspace*{0.015truein}
\centerline{\footnotesize\it Center for Particle Physics, University of Texas}
\baselineskip=10pt
\centerline{\footnotesize\it Austin, Texas 78712, U.S.A.}
\vspace*{10pt}
\centerline{\footnotesize WAI-YEE KEUNG}
\vspace*{0.015truein}
\centerline{\footnotesize\it Physics Department, University of Illinois}
\baselineskip=10pt
\centerline{\footnotesize\it Chicago, Illinois 60607}
\vspace*{10pt}
\centerline{\footnotesize TZU CHIANG YUAN}
\vspace*{0.015truein}
\centerline{\footnotesize\it Davis Institute for High Energy Physics, 
University of California}
\baselineskip=10pt
\centerline{\footnotesize\it Davis, California 95616}
\vspace*{0.225truein}
\publisher{(received date)}{(revised date)}

\vspace*{0.21truein}
\abstracts{The conventional color-singlet model was challenged by the recent
data on quarkonium production.  Discrepancies 
in production rates were observed 
at the Tevatron, at LEP, and in fixed-target experiments.
The newly advocated color-octet mechanism provides a plausible 
solution to the anomalous quarkonium production observed at the Tevatron.  
The color-octet mechanism should also affect other
quarkonium production channels.  
In this paper we will summarize the studies of quarkonium
production in $Z^0$ and $\Upsilon$ decays.
}{}{}

\setcounter{footnote}{0}
\renewcommand{\thefootnote}{\alph{footnote}}

%\textlineskip			%) USE THIS MEASUREMENT WHEN THERE IS
%\vspace*{12pt}			%) NO SECTION HEADING

\vspace*{1pt}\textlineskip	%) USE THIS MEASUREMENT WHEN THERE IS
\section{Introduction}	%) A SECTION HEADING
\vspace*{-0.5pt}

Charmonium was first discovered more than twenty years ago, which was 
believed to be a pair of charm and anti-charm quarks bounded by strong
QCD forces.  The production and decay of quarkonia had then become very
interesting subjects in QCD.  The production of quarkonia was first formulated
by the color evaporation model\cite{evapor}.  
It states that the production rate of a 
particular quarkonium state $H$ is a certain fraction $f_H$ of the open charm
pair production with invariant mass $m_{c\bar c}$ between $2m_c$ and $2m_D$,
i.e., for example,
\begin{equation}
\sigma(\psi) = f_\psi  \times \int^{2m_D}_{2m_c} \frac{\sigma(c\bar c)}
{dm_{c\bar c}} \; dm_{c\bar c} \; .
\end{equation}
The model has the predictive power that $f_H$'s are process independent, and
can reproduce the $\sqrt{s}$ dependence of the total cross section.
However, it is unable to predict the relative rates for different 
spin-orbital states.

The color evaporation model was later superseded by the color-singlet model
(CSM)\cite{csm}.  This model is based on the assumption that the 
spin-orbital angular momentum and the color of the asymptotic physical state
is the same as the point-like $c\bar c$ pair produced in the short distance
region.  Thus, the amplitude involving a physical charmonium state is 
proportional to the amplitude involving the point-like $c\bar c$ pair in the 
same angular momentum state with $c\bar c$ in a color-singlet state and 
having small
relative momentum.  The proportionality constant is given by the wavefunctions
or derivatives of wavefunctions evaluated at the origin, which are obtained 
by solving the Schr\"{o}indger equation of the $c$ and $\bar c$
system.  
%The CSM has high predictive power because it needs only a few
%universal parameters.

The CSM is simple  with high predictive power. However, it
often gives quarkonium production rates below those measured from the
fixed-target experiments. 
%
%has been very successful in describing experimental data until the 
%measurements of prompt $\psi$ and $\psi'$ production by CDF\cite{cdf} at the
%Tevatron a few years ago.  The prediction by the CSM is orders of magnitude
%
The contrast becomes much prominent when measured rates of prompt
$\psi$ and $\psi'$ in CDF experiment\cite{cdf} exceed those
predicted by the CSM by orders of magnitude 
at the large transverse momentum region.  The inadequacy of 
CSM led to two important theoretical developments in quarkonium 
production\cite{annrev}.
The first one is the idea of gluon\cite{gfrag} and heavy quark 
fragmentation\cite{cfrag}.  Though
these fragmentation processes are higher order corrections in $\alpha_s$,
they have been shown\cite{bdfm} to be more important than the lowest order 
QCD production\cite{fusion}
in the large $p_T$ region.  The dominant contribution to $\psi$ production
would be gluon fragmentation $g^*\to \chi_{cJ}$ followed by the radiative 
decays $\chi_{cJ}\to \psi \gamma$.  For $\psi'$ the dominant contributions 
come from gluon and charm quark fragmentation.  However, the
$\psi$ data showed that the $\psi$ fed down from $\chi_{cJ}$ decay
is not the major source of $\psi$\cite{cdf}.  
Moreover, the $\psi'$ data was still 
orders of magnitude above the prediction even including fragmentation
contributions.  

The second important development is the color-octet 
mechanism\cite{braaten-fleming}, which includes
higher order corrections in $v^2$, where $v$ is the relative velocity of $c$
and $\bar c$ inside the charmonium.  Within Non-Relativistic QCD 
(NRQCD)\cite{BBL}
formalism of the quarkonium, the physical state, e.g. $\psi$, is not solely a 
$c\bar c$ pair in $^3S_1$ color-singlet state, instead, a superposition of 
all Fock states:
\begin{eqnarray}
\vert \psi \rangle &=& \vert c\bar c(^3S_1, {\underline 1} )\rangle + 
{\cal O}(v) \;\vert c\bar c (^3P_J, {\underline 8})g \rangle 
+ {\cal O}(v^2)  \vert c\bar c (^3S_1, \underline{8} \; {\rm or} \; 
                   \underline{1} ) gg \rangle 
\nonumber \\
&+& {\cal O}(v^2)  \vert c\bar c(^1S_0,\underline{8})g \rangle 
 + {\cal O}(v^2) \; \vert c\bar c(^3D_{J'}, \underline{8} \; {\rm or} \; 
\underline{1} )gg \rangle + {\cal O}(v^3) \; ,
\label{fock-psi}
\end{eqnarray}
where the angular momenta of the $c\bar c$ pair in each Fock state is labeled
by $^{2S+1}L_J$ with a color configuration of either $\underline{8}$ or
$\underline{1}$.  Similar expressions exist for $\chi_{cJ}$ states. 
Production of $\psi$ can now be via higher Fock states with emission or
absorption of dynamical gluons.  The NRQCD factorization formula for the 
production of a quarkonium state $H$ is given by
\begin{equation}
\label{factor}
d \sigma (H + X)  =
\sum_n d \hat \sigma (c \bar c (n) + X)  
\langle {\cal O}^H_n \rangle \; ,
\end{equation}
where $d\hat\sigma$ is the short distance factor producing the $c\bar c$
pair in an intermediate state $n$ ($n$ is a collective notation for angular
momentum and color), and $\langle {\cal O}^H_n \rangle$ is the matrix 
element describing the transition of the state $n$ into the physical state 
$H$.  The relative importance of each term in (\ref{factor}) depends on the
order of $\alpha_s$ in $d\hat \sigma$ and $v^2$ in 
$\langle {\cal O}^H_n \rangle$.  The CSM is essentially the leading term
in the expansion (\ref{factor}).
The best known example\cite{braaten-fleming,CGMP,cho-leibovich} 
of color-octet mechanism is the gluon fragmentation
into a $c\bar c(^3S_1,\underline{8})$ intermediate state, which then evolves
nonperturbatively into the physical $\psi$ or $\psi'$.  This is so far the
best plausible solution to the anomalous $\psi$ and $\psi'$ production
at the large $p_T$ region in hadronic collisions.  

If the color-octet mechanism is correct, it would have significant impact
on other production modes, e.g., photoproduction\cite{photo}, 
hadroproduction\cite{pN}, production
in $Z^0$\cite{Z0,cho}, $\Upsilon$\cite{upsilon}, and $B$ decays\cite{Bdecay}, 
and at $e^+e^-$ collisions\cite{ee}.  While 
other modes are summarized by other speakers, we concentrate on the 
production in $Z^0$\cite{Z0} and $\Upsilon$\cite{upsilon} decays.

\section{$\psi$ Production in $Z^0$ Decay\cite{Z0}}
\vspace*{-0.5pt}
\noindent
Though the major source of $\psi$ in $Z^0$ decay comes from $B$ meson decays, 
the LEP detectors are able to separate the prompt $\psi$ from the
$\psi$ coming from $B$ decay.  The recent 
data on prompt $\psi$ production showed
a factor of almost 10 larger than the prediction by the CSM.  Therefore, it
would be interesting to examine the effects of the color-octet mechanism in 
this production mode.

Let us first summarize a few important color-singlet processes.
The leading order color-singlet processes are 
$Z \to \psi g g$\cite{keung} and $Z \to \psi c \bar c$\cite{bck}. 
Although both processes are of order $\alpha_s^2$, the latter is two orders 
of magnitude larger. The latter process has been interpreted as a 
fragmentation process\cite{cfrag} 
in which the $c \bar c$ pair was first produced 
almost on-shell from the $Z$ decay and then followed by the fragmentation of 
$c$ or $\bar c$ into the $\psi$\cite{cfrag}. 
Therefore, the process $Z\to \psi c \bar c$ is not suppressed by the quark
propagator effect of order $1/M_Z$ as it does in the process 
$Z\to \psi gg$.  
In the fragmentation approximation, the energy distribution of $\psi$ in 
the process $Z\to \psi c\bar c$ is given by\cite{cfrag}
\begin{equation}
\label{14}
\frac{d\Gamma}{dz}\left(Z\to \psi(z) \, c\bar c  \right) \approx 
2\; \Gamma(Z\to c\bar c) \times D_{c\to \psi}(z) \;,
\end{equation}
where 
\begin{equation}
D_{c\to \psi}(z) = \frac{16\alpha_s^2(2m_c) \langle {\cal O}_1^{\psi}(^3S_1)
\rangle }{243 m_c^3} \frac{z(1-z)^2}{(2-z)^6} \; 
\left(16-32z+72z^2-32z^3+5z^4 \right ) \;,
\end{equation}
and $z=2E_\psi/M_Z$ with $E_\psi$ denotes the energy of $\psi$. 
Both processes involve the same matrix element 
$\langle {\cal O}_1^\psi (^3S_1) \rangle$ whose value can be extracted from 
the leptonic width to be about $0.73 \; {\rm GeV}^3$.  Numerically,
the widths for $Z\to \psi gg$ and $Z \to \psi c \bar c$ are 
about $6\times 10^{-7}$ GeV and $7\times 10^{-5}$ GeV, respectively. 
There is also a higher order color-singlet process $Z\to q \bar q g^*$ 
followed by the gluon fragmentation\cite{hagi} $g^* \to \psi gg$.
The branching ratio was estimated to be of order $10^{-6}$ 
only\cite{gfrag,hagi}.
However, the recent results from LEP\cite{LEP} showed a branching ratio 
of order $10^{-4}$ for prompt $\psi$ production, which is 
well above all the predictions from the color-singlet model. 

\begin{figure}[t]
\centering
\leavevmode
\epsfysize=60pt
\vspace*{8pt}
\epsfbox{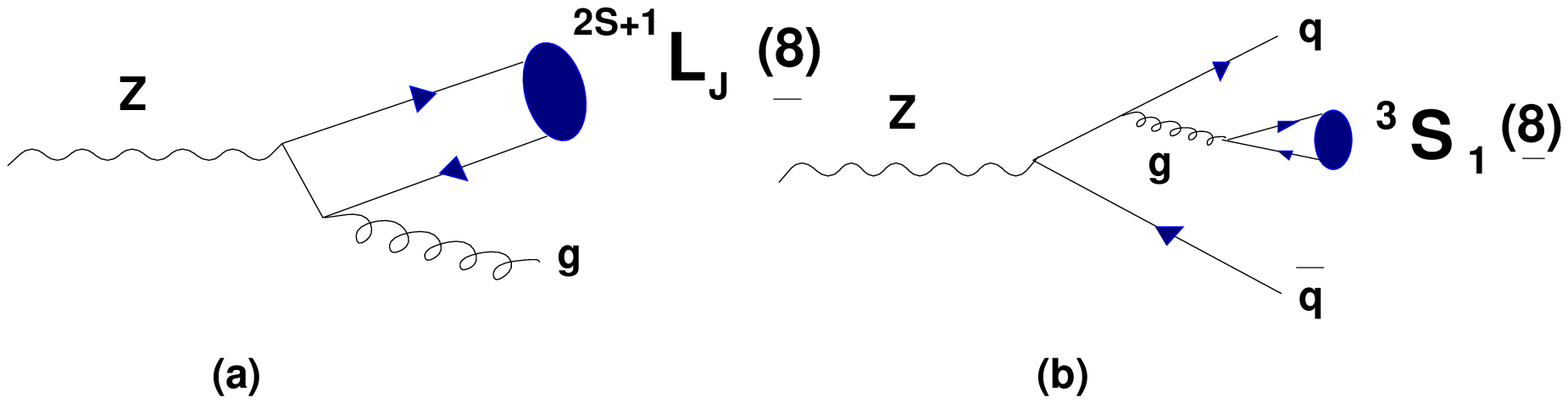}
\vspace*{8pt}
\fcaption{\label{fig1}
Representative Feynman diagrams for (a)$Z\to \psi g$, and
(b) $Z\to \psi q \bar q$.}
\end{figure}

The leading order color-octet process is of order $\alpha_s$ given by 
the process $Z \to \psi g$, for which one of the Feynman diagrams is shown 
in Fig.~\ref{fig1}(a). 
Unfortunately, the short-distance factors are suppressed by at least one 
power of $1/M_Z$. Numerically, the width for $Z\to \psi g$ is of 
order $10^{-7}$ GeV which renders this process useless at the $Z$ resonance. 
  
The dominant color-octet process actually begins at the order $\alpha_s^2$ 
in the process $Z \to \psi q \bar q$, for which one of the Feynman diagrams
is shown in Fig.~\ref{fig1}(b). Here $q$ represents $u,d,s,c$, or $b$. 
The energy distribution of $\psi$ for the process $Z \to \psi q \bar q$ 
is calculated, in the limit $m_q = 0$, to be\cite{Z0}
\begin{eqnarray}
\label{exact}
\frac{d \Gamma}{d z}(Z  \to \psi (z) q \bar q) 
&=&  \frac{\alpha_s^2(2m_c)}{18} \Gamma (Z \to q \bar q) 
\frac{\langle O_8^{\psi} (^3S_1) \rangle }{m_c^3}  \\
&\times & \left[ \left( \frac{(z-1)^2 + 1}{z} + 2 \xi \frac{2 - z}{z} + 
\xi^2 \frac{2}{z} \right) \log \left( \frac{z + z_L}{z-z_L} \right) 
- 2 z_L \right] \; , \nonumber
\end{eqnarray}
where $z=2E_\psi/M_Z$, $\xi=M_\psi^2/M_Z^2$, and $z_L = (z^2 - 4 \xi)^{1/2}$. 
The physical range of $z$ is $2\sqrt{\xi} <z< 1+\xi$. 
Integrating Eq.~(\ref{exact}) over the physical range of $z$ gives 
the decay width of $Z \to \psi q \bar q$\cite{Z0}
\begin{eqnarray}
&&
{\Gamma (Z \to \psi q\bar q)\over \Gamma(Z\to q\bar q)}=
\frac{\alpha_s^2(2m_c)}{36} \frac{\langle O_8^{\psi} (^3S_1) \rangle }{m_c^3} 
\bigg\{5(1-\xi^2)-2\xi\log\xi+ (1+\xi)^2  \times \\
&&\quad \bigg[2\,\hbox{Li}_2\left({\xi\over 1+\xi}\right) 
            - 2\,\hbox{Li}_2\left({ 1 \over 1+\xi}\right)
     -2\log(1+\xi)\log\xi + 3\log\xi+\log^2\xi\bigg] \bigg\} \; ,
\nonumber
\end{eqnarray}
where Li$_2(x)=-\int_0^x{dt\over t}\log(1-t)$ is the Spence function.
Numerically, the above ratio is $2.2\times 10^{-4}$.
Summing over all the quark flavors $(q=u,d,s,c,b)$, we obtain 
the decay width 
$\sum_q \Gamma(Z\to \psi q\bar q) \simeq 3.8\times 10^{-4}$ GeV 
and the branching ratio 
$\sum_q {\rm Br}(Z\to \psi q\bar q) \simeq 1.5 \times 10^{-4}$.
Assuming the dominant prompt $\psi$ production process to be 
$Z\to q\bar q g^*$ with $g^* \to \psi+X$, in which according to color-singlet
model the off-shell gluon fragments into a $\psi$ plus two perturbative 
gluons, DELPHI\cite{LEP} obtained the limit 
${\rm Br}(Z\to q \bar q g^*; \; g^* \to \psi +X) 
< 4.1 \times 10^{-4}$.
Since the event topology of our color-octet process is similar to this one, 
this limit should also be valid for our color-octet process.  Therefore,
our result is consistent with this data.

\begin{figure}[t]
\centering
\leavevmode
\epsfysize=180pt
\epsfbox{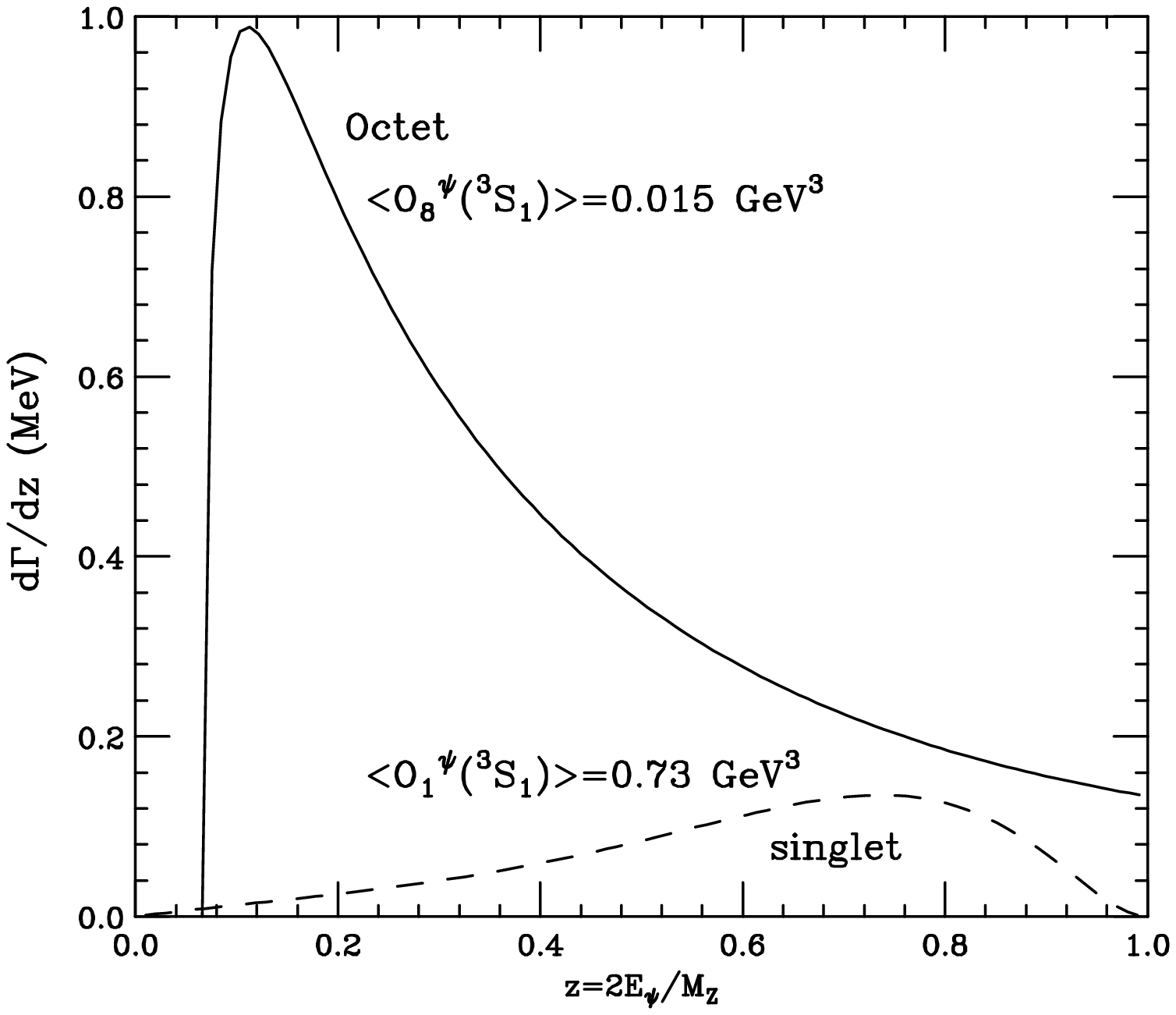}
\vspace*{12pt}
\fcaption{\label{fig2}
Energy spectrum $d\Gamma/dz$ of $\psi$ from the color-octet process
$Z\to \psi q\bar q$ and the color-singlet process $Z\to \psi c\bar c$.}
\end{figure}

In Fig.~\ref{fig2}
 we compare the energy distributions of $\psi$ coming from the most
important color-octet process $Z\to \psi q\bar q$ using Eq.~(\ref{exact}) 
and the most dominant color-singlet process $Z \to \psi c \bar c$ 
using  Eq.~(\ref{14}). 
The comparison in Fig.~\ref{fig2}
 shows a very spectacular difference between the 
color-octet and color-singlet processes.
The spectrum for the color-octet process is very soft with a pronounced peak 
at the lower $z$ end, while the spectrum for the color-singlet process is
rather hard because of the nature of heavy quark fragmentation\cite{cfrag}. 

Another possible test to distinguish the charm quark fragmentation 
$Z\to \psi c\bar c$ and the color-octet process $Z\to \psi q \bar q$ is 
using the polarization of $\psi$.  The polarization of $\psi$ can be measured
by the angular distribution of the lepton pair in the 
rest frame of $\psi$.  The charm quark fragmentation process predicts a 
polarization ratio $L/(L+T) \approx 31\%$, while the dominant color-octet 
process gives a ratio $L/(L+T)\approx 20\%$.  Nevertheless, the small
production rate of prompt $\psi$ renders this test rather difficult.

Concluding this section the color-octet process $Z\to \psi q \bar q$ 
contributes dominantly to the prompt $\psi$ production, and the resulting 
branching ratio is consistent with the data.  In addition, the energy spectrum
of $\psi$ serves as a useful test to distinguish between the color-octet 
process $Z\to \psi q\bar q$ and the color-singlet process $Z\to \psi c\bar c$.

\section{$\psi$ Production in $\Upsilon$ Decay\cite{upsilon}}
\vspace*{-0.5pt}
\noindent
The available experimental data on charmonium production in 
$\Upsilon$ decay are  listed as follows,
\begin{equation}
   \hbox{Br} (\Upsilon \to \psi + X)
   \ \left\{
   \begin{array}{ll}
   = (1.1 \pm 0.4) \times 10^{-3} 
       & \mbox{{\small CLEO}\cite{CLEO},}    \\
   <  1.7 \times 10^{-3}
       & \mbox{Crystal Ball\cite{Crystal},}  \\ 
   <  0.68 \times 10^{-3}
       & \mbox{{\small ARGUS}\cite{ARGUS}.}
   \end{array}
   \right.
\label{Bexp}
\end{equation}
Such large branching ratios by CLEO can 
hardly be explained by the CSM  
because the lowest order color-singlet processes are of order $\alpha_s^6$
and there is no quantitative calculation of such existing in the literature.
Only until a few years ago an indirect $\psi$ production 
via the intermediate physical $\chi_c$ states was performed
by Trottier\cite{trottier}.  This process is of order $\alpha_s^5$, i.e.,
$1/\alpha_s$ larger than the color-singlet processes.
Nevertheless, the branching ratio obtained is still one order of magnitude
below the data. 
Therefore, it would be interesting to identify important 
color-octet processes.  

The NRQCD factorization formula\cite{BBL} for the  inclusive charmonium 
production in bottomonium decay, such as 
$\Upsilon \to \psi + X$, is given by 
\begin{equation}
\label{double_fact}
d \Gamma (\Upsilon \to \psi + X) = 
\sum_{m,n} d \widehat \Gamma_{mn} 
\langle \Upsilon \vert O_m \vert \Upsilon \rangle 
\langle  O_n^\psi  \rangle \; ,
\end{equation}
where $d \widehat \Gamma_{mn}$ are the short-distance factors for a $b \bar b$ 
pair in the state $m$ to decay into a $c \bar c$ pair in the state $n$ plus 
anything, where
$m,n$ denote collectively the color, total spin, and orbital 
angular-momentum of the heavy quark pair.  
$d \widehat \Gamma_{mn}$ can be calculated in perturbation theory as 
a series expansion in $\alpha_s(m_c)$ and/or $\alpha_s(m_b)$.
The nonperturbative factor 
$\langle \Upsilon \vert O_m \vert \Upsilon \rangle$ is proportional 
to the probability for the $b \bar b$ pair to be in the state $m$ inside the 
physical bound state $\Upsilon$, while 
$\langle  O_n^\psi  \rangle$ is proportional to 
the probability for a 
point-like $c \bar c$ pair in the state $n$ to form the bound state $\psi$. 
The relative importance of the various terms in the above double 
factorization formula (\ref{double_fact}) can be determined by the order 
of $v_{b}$ or $v_c$ in the NRQCD 
matrix elements and the order of $\alpha_s$ in the 
short-distance factors $d \widehat \Gamma_{mn}$. 
In fact, the calculation by Trottier\cite{trottier} is an example of this
factorization formula.
An infrared divergence occurs in the leading order calculation  
of the short-distance factor,
therefore, in addition to the color-singlet matrix element 
$\langle  O_1^{\chi_{cJ}}(^3P_J)  \rangle$, one also needs to 
include the color-octet matrix element 
$\langle  O_8^{\chi_{cJ}}(^3S_1)  \rangle$ 
to absorb the infrared divergence.  
In this case, the introduction of the 
color-octet matrix element is required by perturbative consistency.

For color-octet processes we first consider the produced 
$c \bar c$ pair in the color-octet 
$^1S_0$, $^3S_1$, or $^3P_J$ configuration, which subsequently evolves 
into physical $\psi$ described by the matrix elements 
$\langle  O^\psi_8 (^1S_0)  \rangle$, 
$\langle  O^\psi_8 (^3S_1)  \rangle$, or  
$\langle  O^\psi_8 (^3P_J)  \rangle$, respectively.
The matrix element 
$\langle  O^\psi_8 (^3S_1)  \rangle$ has been extracted from the 
CDF data\cite{braaten-fleming,CGMP,cho-leibovich}, while 
two different combinations of the other color-octet matrix elements have   
been extracted from the CDF data\cite{cho-leibovich} and
{}from the photoproduction data by Amundson {\it et al}\cite{photo}.  
Though of much smaller effects, we also consider the contributions by the
higher Fock state of the color-octet $b \bar b$ pair 
inside the $\Upsilon$ associated with the matrix element 
$\langle \Upsilon \vert O_8 (^3S_1) \vert \Upsilon \rangle$, 
whose value has not yet been determined.  
An order of magnitude of this matrix element can, in principle, be estimated
by considering the ratio
\begin{equation}
\label{ratio1}
\frac{\langle \Upsilon |O_8(^3S_1) |\Upsilon \rangle }
     {\langle \Upsilon |O_1(^3S_1) |\Upsilon \rangle } \;
\frac{\langle  O_1^\psi(^3S_1)  \rangle }
     {\langle  O_8^\psi(^3S_1)  \rangle } \sim
\left( \frac{v_b^2}{v_c^2} \right)^2 \;, 
\end{equation}
which implies $\langle \Upsilon |O_8(^3S_1) |\Upsilon \rangle \approx 
3 \times 10^{-3}$ GeV$^3$.  
The ratio in (\ref{ratio1}) tells us that processes
associated with a color-octet $c \bar c$ pair inside the produced $\psi$ 
and a color-singlet $b \bar b$ pair inside the decaying $\Upsilon$ 
are much more important than
those with a color-octet $b \bar b$ pair inside the $\Upsilon$ and a 
color-singlet $c \bar c$ pair inside the $\psi$. 
However, such a large value for this matrix element would substantially
increase the hadronic width of $\Upsilon$, which would diminish the 
leptonic branching ratio to an unacceptable level.
In order not to spoil the experimental value for the leptonic branching ratio
and the total hadronic width of $\Upsilon$, we necessarily obtain
the following bound on $\langle \Upsilon |O_8(^3S_1) |\Upsilon \rangle$: 
\begin{equation}
\label{bound}
\langle \Upsilon |O_8(^3S_1) |\Upsilon \rangle \approx
\left(1.9 {  \begin{array}{c}  +5.1\\ -4.6 \end{array}  } \right) 
\times 10^{-4} \; {\rm GeV}^3  \ .
\end{equation} 
Therefore, the color-octet processes associated with 
$\langle \Upsilon |O_8(^3S_1) |\Upsilon \rangle$ have very small branching 
ratios, and thus negligible.

\begin{figure}[t]
\centering
\leavevmode
\epsfysize=150pt
\epsfbox{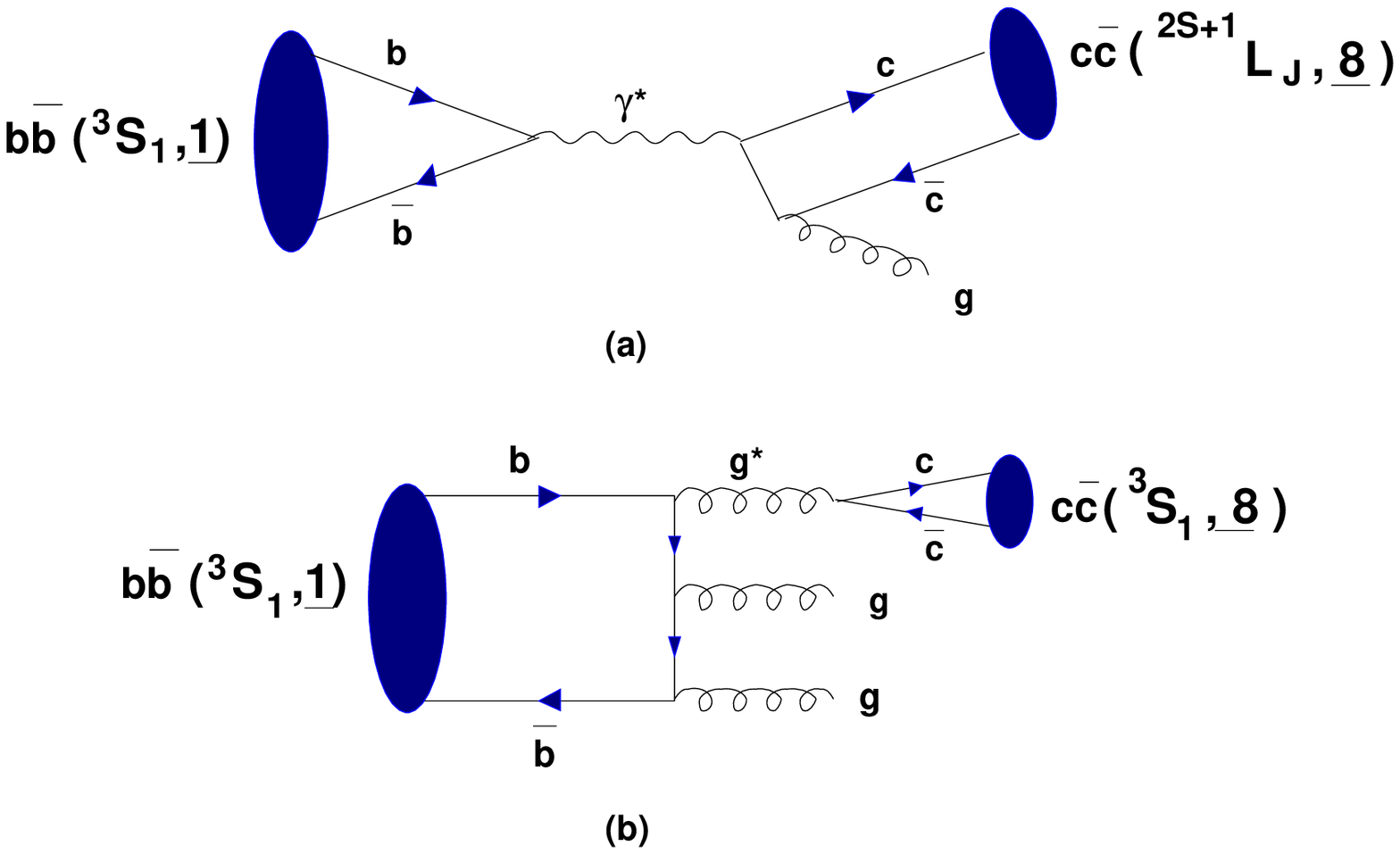}
\vspace*{12pt}
\fcaption{ \label{fig3}
Some of the contributing Feynman diagrams for the short-distance 
processes:
(a) $b \bar b (^3S_1, \underline{1}) \to \gamma^*   \to 
          c \bar c (^{2S+1}L_{J}, \underline{8})  g$;
(b) $b \bar b (^3S_1, \underline{1}) \to g^*gg \to   
          c \bar c (^3S_1, \underline{8})  gg$;
}
\end{figure}

The first color-octet process we consider is the one with the short distance
process $b \bar b (^3S_1,\underline{1}) \to c \bar c 
(^{2S+1}L_J,\underline{8}) +g$, which is of order $\alpha^2 \alpha_s$.
One of the Feynman diagrams is shown in Fig.~\ref{fig3}(a).
Using the heavy quark spin symmetry relation $\langle O_8^\psi(^3P_J) \rangle 
\approx (2J+1) \langle O_8^\psi(^3P_0) \rangle$\cite{BBL}, 
the total width from these processes can be simplified as 
\begin{eqnarray}
\Gamma_{a}(\Upsilon \to \psi + X) &=& \frac{4 \pi^2 Q_c^2 Q_b^2 \alpha^2 
\alpha_s}{3} \; 
\frac{\langle \Upsilon \vert O_1(^3S_1) \vert \Upsilon \rangle}
{m_b^4m_c} \;
\biggr\{ \langle  O_8^\psi(^1S_0)  \rangle (1-\xi) 
\\
&+&  
 \frac{\langle  O_8^\psi(^3P_0)  \rangle }{3 m_c^2} \biggr[
\frac{(1-3\xi)^2}{1-\xi} +\frac{6(1+\xi)}{1-\xi} +
\frac{2(1+3\xi+6\xi^2)}{1-\xi} \biggr] \biggr\} \ . \nonumber 
\end{eqnarray}
Using\cite{cho-leibovich} $\langle  O_8^\psi (^1S_0)  \rangle 
\approx \langle  O_8^\psi (^3P_0)  
\rangle /m_c^2 \approx 10^{-2}\, {\rm GeV}^3$, 
the contribution from the above color-octet processes 
to the inclusive branching ratio BR$(\Upsilon \to \psi +X)$ is only 
$1.6 \times 10^{-5}$, which is 
almost two orders of magnitude below the CLEO data (\ref{Bexp}).

The next process we consider is $b \bar b (^3S_1, \underline{1}) \to ggg^*
\to c \bar c (^3S_1, \underline{8}) + g g$, which is of order $\alpha_s^4$.
Fig.~\ref{fig3}(b) shows one of the six Feynman diagrams for the 
$b \bar b (^3S_1, \underline{1})$ 
pair annihilating into three gluons with one of the gluons converting into the 
$c \bar c (^3S_1, \underline{8})$ pair. 
The partial width is this process is given by\cite{upsilon}
\begin{equation}
\label{width1}
\Gamma_{b}(\Upsilon \to \psi +X) = 
\frac{5 \pi \alpha_s^4}{486 m_c^3 m_b^2}
\langle \Upsilon \vert O_1(^3S_1) \vert  \Upsilon \rangle 
\langle  O_8^\psi(^3S_1)  \rangle
 \; \times (0.571) \; .
\end{equation}
We note that the color-octet piece for the process 
$\Upsilon \to \chi_{cJ} + X$ considered by Trottier\cite{trottier} 
can be obtained from the above Eq.~(\ref{width1})  by simply replacing 
the matrix element 
$\langle  O_8^\psi(^3S_1)  \rangle$ with
$\langle  O_8^{\chi_{cJ}}(^3S_1)  \rangle$. 
The prediction of the partial width is sensitive to the values  
of the two NRQCD matrix elements, the running coupling constant  
$\alpha_s$, and the heavy quark masses. 
For convenience we can normalize this partial width to the three-gluon width 
$\Gamma(\Upsilon \to ggg)$
\begin{equation}
\label{ratio}
\frac{\Gamma_{b}(\Upsilon \to \psi +X)}{\Gamma(\Upsilon \to ggg)}
= \frac{\pi\alpha_s}{8}  
\frac{\langle  O^\psi_8(^3S_1)  \rangle}{m_c^3}  
\;\frac{0.571}{\pi^2 - 9}\; .
\end{equation}
With $\langle  O_8^\psi(^3S_1) \rangle = 0.014 \; 
{\rm GeV}^3$\cite{braaten-fleming,cho-leibovich}, and assuming 
BR$(\Upsilon \to ggg)  \approx$ BR$(\Upsilon \to {\rm light \; hadrons}) 
= 0.92$\cite{pdg}, we obtain 
\begin{eqnarray}
\frac{\Gamma_{b}(\Upsilon \to \psi +X)}{\Gamma_{\rm total}(\Upsilon)} 
& = &
\frac{\Gamma_{b}(\Upsilon \to \psi +X)}{\Gamma(\Upsilon \to ggg)}
\; \times \; {\rm BR}(\Upsilon \to ggg) 
\nonumber \\
& \approx & 2.5 \times 10^{-4} \; \; .
\end{eqnarray}
This  prediction is smaller than the CLEO data by merely a factor of 4, 
and is consistent with the bounds from Crystal Ball and ARGUS.

The above color-octet process can be extended to $\psi'$ as well, 
simply by replacing the matrix element 
$\langle O_8^\psi (^3S_1) \rangle$ with 
$\langle O_8^{\psi'} (^3S_1) \rangle$, whose 
value has also been extracted from the CDF data to be 0.0042 
GeV$^3$\cite{braaten-fleming,CGMP,cho-leibovich}. 
With\cite{pdg} BR$(\psi' \to \psi + X) \approx 57 \%$, 
we obtain a branching 
ratio of $4.3 \times 10^{-5}$ for the inclusive $\psi$ production in the 
$\Upsilon$ decay that comes indirectly from $\psi'$. 
One can also consider the processes 
$\Upsilon \to \gamma g g^*$ followed by $g^* \to \psi \; (\psi')$ 
via the color-octet mechanism, and 
$\Upsilon \to gg \gamma^*$ followed by $\gamma^* \to \psi \; (\psi')$ 
in the color-singlet model.    Their partial widths  
are suppressed by factors of $8 \alpha/(15 \alpha_s) \sim 0.02$  and 
$32 \alpha^2 \langle O^\psi_1(^3S_1) \rangle
/(45 \alpha_s^2 \langle O^\psi_8(^3S_1) \rangle) \sim 0.06$, respectively,
compared with the width of Eq.(\ref{width1}). Thus they 
contribute a branching fraction about $2 \times 10^{-5}$ in the inclusive 
decay $\Upsilon \to \psi +X$. The indirect contribution from the $\psi'$ 
{}from these two processes is about $6\times 10^{-6}$.

Another process we consider is  that the $b\bar b$ pair annihilates into a 
$q\bar q$ pair via the $s$-channel photon 
$\gamma^\ast$, and then a bremsstrahlung virtual gluon emitted from the light 
quark line splits into an octet $c\bar c(^3S_1,\underline{8})$, 
which then turns into $\psi$. 
This process is of order $\alpha^2\alpha_s^2$, which is similar to the
 dominant color-octet process in the prompt $\psi$ production in 
$Z^0$ decay\cite{Z0}. 
The partial width is given by  
\begin{eqnarray}
\label{e:dl}
{\Gamma_{1e} (\Upsilon\to \gamma^* \to q \bar q \psi)
\over 
\Gamma(\Upsilon \to \gamma^* \to q\bar q)}    
&=&
\frac{\alpha_s^2(2m_c)}{36} \frac{\langle O_8^{\psi} (^3S_1) \rangle }{m_c^3} 
\bigg\{5(1-\xi^2)-2\xi\log\xi  \nonumber \\
&+& \bigg[
 2\,\hbox{Li}_2\left({\xi\over 1+\xi} \right)
 -  2\,\hbox{Li}_2\left({ 1 \over 1+\xi}\right)
-2\log(1+\xi)\log\xi \nonumber \\
&& + 3\log\xi+\log^2\xi\bigg](1+\xi)^2\bigg\} \, ,
\end{eqnarray}
with $\xi = (m_c/m_b)^2$. 
Numerically, the ratio in (\ref{e:dl}) is only $8.3 \times 10^{-6}$, and so
this process can be safely ignored.

Among all contributions, the largest comes from the process shown in 
Fig.~\ref{fig3}(b),
which has a branching ratio about $2.5 \times 10^{-4}$. 
The next largest contribution is the indirect process considered by 
Trottier\cite{trottier}, which  gives a branching ratio of 
$5.7 \times 10^{-5}$.  
Other comparable processes are 
(1) the process shown in Fig.~\ref{fig3}(a) which has a branching ratio 
of $1.6 \times 10^{-5}$, 
(2) the processes $\Upsilon \to gg \gamma^* \to \psi +X$ and 
$\Upsilon \to \gamma g g^* \to \psi + X$ via the color-octet mechanism 
which have a combined branching ratio about $2 \times 10^{-5}$, 
and finally, (3) the indirect contribution from $\psi'$ 
having a branching ratio about $5 \times 10^{-5}$. 
Therefore, adding up the contributions from all these processes,  
we obtain a branching ratio BR$(\Upsilon \to \psi + X)  \approx 
4 \times 10^{-4}$, which is within $2\sigma$ of the CLEO data.  

\section{Conclusions}
\vspace*{-0.5pt}
\noindent
The color-singlet model, which has been popularly used for more than 15 years,
is inadequate to describe the present data at the Tevatron, LEP,
$\Upsilon$ decay, and fixed target experiments.  The NRQCD description of 
quarkonium production provides a consistent theoretical framework and 
implies the so-called color-octet mechanism.  
It can explain the large $p_T$ $\psi$ and 
$\psi'$ data at the Tevatron by fitting a small set of parameters of matrix 
elements. If the color-octet
mechanism is correct, it would have a strong impact on other production modes.
In this talk, we summarized the $\psi$ production in $Z^0$ and $\Upsilon$
decays. The color-octet mechanism turns out to be very important in these
modes. 
The color-octet process $Z\to \psi q\bar q$ 
can explain the data on $\psi$ production in 
$Z^0$ decay.  A more decisive test would be a measurement 
of the energy profile of the $\psi$ where a very soft spectrum 
is predicted by the color-octet mechanism. 
Furthermore, in $\Upsilon$ decay the color-octet mechanism makes possible a new
short distance process $b\bar (^3S_1,\underline{1}) \to ggg^* \to gg c\bar c
(^3S_1,\underline{8})$, which can bring the theoretical prediction
to be within $2\sigma$ of the central value of the CLEO data.

\nonumsection{Acknowledgements}
\noindent
This work was supported in part by the United States Department of 
Energy under Grant Numbers DE-FG02-84ER40173, DE-FG03-93ER40757,  
and DE-FG03-91ER40674.

\nonumsection{References}

\end{document}